# Linear complexions: Metastable phase formation and coexistence at dislocations

Vladyslav Turlo [a], Timothy J. Rupert [a, b, *]

[a] Department of Mechanical and Aerospace Engineering, University of California, Irvine, CA 92697, USA

[b] Department of Chemical Engineering and Materials Science, University of California, Irvine, CA 92697, USA

* Corresponding Author: trupert@uci.edu

**Abstract**

The unique three-phase coexistence of metastable B2-FeNi with stable $L1_0$-FeNi and $L1_2$-FeNi$_3$ is discovered near edge dislocations in body-centered cubic Fe-Ni alloys using atomistic simulations. Stable nanoscale precipitate arrays, formed along the compression side of dislocation lines and defined as *linear complexions*, were observed for a wide range of compositions and temperatures. By analyzing the thermodynamics associated with these phase transitions, we are able to explain the metastable phase formation and coexistence, in the process defining new research avenues for theoretical and experimental investigations.

Solute segregation to crystal defects such as grain boundaries or dislocations can cause the formation of thermodynamically-stable, phase-like states that can be called *complexions* [1-3]. Over the last two decades, a variety of grain boundary complexions have been discovered and classified [2-13], with their substantial impact on a spectrum of physical properties of polycrystalline materials also being uncovered [14-21]. In contrast, linear complexions, defined as stable chemical and structural states confined at line defects or dislocations, have only been just recently found and represent a new prospective field of study in physics and materials science [22-26]. Linear complexions in the form of stable nanoscale precipitate arrays were first reported by Kuzmina et al. [22] in a body-centered cubic Fe-9at.% Mn alloy. These authors also estimated out that one cubic meter of a strained alloy can contain up to one light year of dislocation lines, meaning there is an important opportunity to control material behavior with linear complexions. Linear complexions allow for the possibility of a tunable alloy microstructure, with the dislocation network providing a possible template for controlled segregation and precipitation. For example, stable nanoscale-size precipitates can act as obstacles for dislocation and grain boundary motion, dramatically improving the thermal stability [27,28] and strength [29,30] of a wide variety of engineering alloys. While local segregation represents one example of a dislocation-driven phase transformation, the local stress field by itself may cause a nanoscale second phase precipitation near the dislocation core. For example, the analytical and phase-field models provided by Levitas [31] and Levitas and Javanbakht [32] explore the nucleation of non-equilibrium, high-pressure phases near dislocation pile-ups. Restricted to the dislocation lines, such stable high-pressure phases could also be considered a type of linear complexion. Thus, dislocations are arguably the most important material defect, as they are the key features whose behavior determines strength, ductility, and many other mechanical properties of materials. The ability to control the nucleation and create a pattern of these features would be a powerful tool for materials design.

Despite the importance of linear complexions, a systematic investigation of nanoscale precipitation on dislocations in relation to composition and temperature is likely too time-consuming for an experimental effort, slowing the discovery of new types of linear complexions. Although phase-field [33,34] and lattice-type [35,36] models can treat complexion transitions, these methods consider a limited number of phases and require many important physical parameters associated with the transition to be defined ahead of time. Moreover, such phenomenological models cannot provide atomic-scale details about the structural transformations associated with linear complexion formation. In contrast, atomistic simulations naturally capture most of the chemical and structural information associated with nanoscale phase transformations. For example, atomistic models have provided great insight into grain boundary complexion transformations such as (1) a "split-kite"-to-"filled-kite" transitions at tilt grain boundaries in Ag-doped Cu [6], (2) ordered-to-disordered grain boundary transitions and nanoscale amorphous intergranular film formation in Zr-doped Cu [37], and (3) destruction of the symmetry at a tilt grain boundary in Ni-doped Mo [38].

In this study, the segregation-induced formation of intermetallic linear complexions in Ni-doped body-centered cubic Fe is investigated via hybrid molecular dynamics (MD) and Monte Carlo (MC) atomistic simulations. Complexions are discovered in the form of nanoscale precipitates appearing along dislocation lines and composed of one, two, or even three intermetallic phases, such as a metastable B2-FeNi, stable $L1_0$-FeNi, and stable $L1_2$-FeNi$_3$. Only edge dislocations are considered and investigated in this work, as they have hydrostatic stress components that can drive substitutional solute segregation. We examine a wide range of chemical compositions and temperatures, which are used to construct a *linear complexion phase diagram*. We also apply equilibrium thermodynamics to explain the existence of the metastable B2-FeNi phase as well as multiphase coexistence in the linear complexions. By providing the first clear nanoscale understanding of the thermodynamics of linear complexion formation, this work provides a path forward for the discovery of new types of dislocation-limited phase transformations.

Hybrid MD/MC atomistic simulations were performed using the Large-scale Atomic/Molecular Massively Parallel Simulator (LAMMPS) software package [39] and an embedded-atom method (EAM) interatomic potential for the Fe-Ni system [40]. An MC step in the variance-constrained semi-grand canonical ensemble [41] was carried out every 100 MD steps of 1 fs each. The simulation was stopped when the energy gradient over the prior 1 ns became less than 1 eV/ns. As shown in Supplemental Fig. S1, this criterion ensures that thermodynamic equilibrium has been reached and no further structural changes are occurring. Atomic structures are visualized with OVITO software [42]. The locations of all dislocations were tracked using the dislocation extraction algorithm (DXA) [43] implemented in OVITO.

The initial simulation cell was constructed with one positive and one negative edge dislocation, formed by removing one half of the atomic plane in the middle of the sample and equilibrating with molecular statics. The simulation cell sizes were 23, 24, and 7.5 nm in the X[111], Y[1-10], and Z[11-2] directions, respectively. To ensure a systematic study, a broad range of chemical compositions (1-20 at.% Ni) and temperatures (400-800 K) were considered. With an increase in bulk composition, solute segregation occurs at the compression side of the dislocations and, eventually, an ordered structure with near-equiatomic composition is formed, as shown in Fig. 1. Due to the smaller atomic radius of Ni atoms, their segregation to the compression side of the dislocations dramatically reduces the local stress, as shown in Figs. 1(a)-(c). An increase in the global Ni composition (Fig. S2) or reduction of the temperature (Fig. 2(b)) promotes the formation of a new phase with local chemical order. Analysis of the local atomic order provided in Fig. 2(c) for the same composition and temperature as shown in Fig. 2(b) demonstrate that a B2-FeNi intermetallic phase has formed, as the atoms are colored red. Further increasing the bulk composition at lower temperatures of 400 and 500 K leads to the formation of the $L1_0$-FeNi phase in the center of this B2-FeNi phase, resulting in their coexistence in Fig. 2(d). At 600 K (Fig. 2(e)), the $L1_2$-FeNi$_3$ intermetallic phase is formed separately from the previous two phases, creating a unique phenomenon of the three-phase coexistence. At higher temperatures and bulk compositions (Fig. 2(f)), the $L1_2$-FeNi$_3$ intermetallic phase dominates and large precipitates of only this phase are formed at the dislocation. Figs. 2(g-j) present perspective views of the representative complexions listed above. It is worth mentioning that only the large $L1_2$ precipitates destroy the original dislocations in Figs. 2(f) and (j), while the nanoscale arrays of precipitates or true linear complexions simply nucleate near the dislocations. The $L1_0$ and $L1_2$ phases with face-centered cubic (fcc) structure preferentially grow along the [11-1] direction instead of following the original dislocation line, thus creating a faceted or kinked dislocation line. To show this more clearly, addition simulations were performed with samples that were 10 times longer in the direction of the dislocation line vector. Fig. 2(k) shows the formation of precipitate arrays in the longer sample with 1 at.% Ni annealed at 400 K (i.e., the same conditions shown in Fig. 2(c)). This image shows that the metastable B2 precipitates grow along the dislocation lines but not perpendicular to them, due to the limited size of the dislocation segregation zone. This finding highlights the decisive role of the dislocations in the stabilizing nanoscale precipitates of the metastable phase, leading to the formation of linear complexions.

Fig. 3(a) compiles our observations for all compositions and temperatures, representing the precipitate composition and fraction of the total simulation cell. For low temperatures and global Ni compositions, the formation of small metastable B2-rich precipitates is observed. With an increase in global Ni composition at 400 and 500 K, the size of the precipitates as well as the fraction of $L1_0$ phase increases. The B2 phase again forms first at 600 K, with a continued increase in Ni composition leading to the $L1_0$ and $L1_2$ transitions occurring together to give the three-phase coexistence (e.g., the purple dots in Fig. 3(a)). At 700 K, the transition from B2 to $L1_2$ is relatively abrupt, while no intermetallic phase precipitation is observed at 800 K. The dashed line in Fig. 3(a) represents the solubility limit curve for bulk Fe-Ni computed for the interatomic potential used here [40]. Fe-Ni alloys with a composition below the solubility limit should be in a solid solution state with no intermetallic phase formation expected. However, the dislocation leads to the nanoscale precipitates comprised mainly of the metastable B2 phase, making a linear complexion. Fe-Ni alloys with a composition above the solubility limit should be in a two-phase coexistence state, but the $L1_0$ intermetallic phase formation is only expected based on the bulk phase diagram. However, in the presence of a dislocation, the nanoscale precipitates in these alloys are made of two or even three intermetallic phases coexisting together, again resulting in a different type of linear complexion. Fig. 3(b) demonstrates the precipitate compositions in relation to predictions from the bulk phase diagram, computed for the interatomic potential used in this work [40]. The metastable B2 phase does not appear on the bulk phase diagram, but B2-rich precipitates are observed at dislocations in our study. These particles have compositions close to but slightly below 50 at.% Ni. Local precipitate compositions of the stable $L1_0$ and $L1_2$ phases are ~50-55 at.% Ni and ~65-75 at.% Ni, respectively, in excellent agreement with the computed bulk phase diagram. However, the average concentration of the dislocation segregation zone is usually below

~20 at.% Ni [26] and cannot by itself explain the formation of intermetallic phases with such a high composition.  Kwiatkowski et al. [25] recently suggested that spinodal decomposition can occur inside of the dislocation segregation zone for the Fe-Mn system.  These authors demonstrated that the solute-rich zones can have a composition near or above the 50 at.% Mn needed to promote second-phase nucleation, leading to the formation of precipitate arrays along the dislocation lines.  Fig. S3 shows a schematic of this concept, as well as a local composition map from our simulation to prove that this phenomenon also occurs here.  Interestingly, the formation of these new phases does not destroy the original defect in our simulations and the growth of the metastable phase in Fe-Ni is restricted to the segregation zone.

The nanoscale size of the precipitates formed along the dislocation lines and their restriction to the near-dislocation region allows us to hypothesize that the interfaces and local stress/strain play a significant role in the nucleation of the intermetallic phases.  Atomistic calculations of the various energy contributions to nucleation were performed, with details provided in the Supplemental Information.   Fig. 4 plots the work associated with intermetallic phase formation as a function of the precipitate radius.  The bcc-like structure of the B2 phase allows for a coherent interface with the bcc solid solution, causing the corresponding free energy barrier for such a transition to be three orders of magnitude lower than that for the bcc to $L1_0$ transition.  Moreover, the exponential dependence of the nucleation rate on this nucleation energy barrier value makes such a difference even more significant (see Supplemental Fig. S7, where an estimated dependence of the nucleation probability is presented as a function of temperature).  In addition, Fig. 4 shows that the stable $L1_0$ phase is less favorable than the metastable B2 phase for a particle radius up to ~2.5 nm (or diameter up to ~5 nm), which is larger than the segregation zone we observed near the dislocation.

For the case of Fe-Ni, we can show that linear complexion transformations inside of the dislocation segregation zone proceed through two kinetic stages. First, the metastable B2-FeNi phase is formed and fills the space of the segregation zone. Then, the stable $L1_0$ phase, which is energetically more favorable but restricted due to a high interfacial energy when abutting the bcc phase, can be formed inside of the B2 precipitate. As shown in Fig. S6(a-b), this B2-to-$L1_0$ transformation has a diffusion-less nature and involves lattice-distortive displacements such as dilatation and shear, leading to a change in the simulation box size and shape. In the locked volume of a dislocation segregation zone, where the volume is restricted by the material surrounding resisting the deformation associated with the phase transition, local strains associated with phase transformation at the nanoscale do not allow for a complete transition from the B2 to $L1_0$. This restriction leads to phase coexistence, as shown in Fig. S6(c) and Fig. S8. Thus, some amount of the metastable B2 phase will remain on the border between the bcc solid solution and the $L1_0$ phase precipitates, as was observed in our atomistic simulations (see, e.g., Figs. 2(d,e)).

Taken as a whole, our work uncovers several barriers for discovering and investigating linear complexions, as well as emphasizes the critical role of atomic-scale modeling approaches in the study of these transformations. For example, it has been recently shown in experiment [23] and in our previous atomistic simulations [26] that linear complexions appear at dislocations in the form of nanoscale precipitate arrays, either because of the spinodal decomposition in the dislocation segregation zone [25] or because of the nucleation and directional growth of precipitates [26]. The two-dimensional visualization methods such as transmission electron microscopy that are typical of experimental investigations commonly used to characterize grain boundary complexions have not been able to capture the repeating structure of linear complexions along dislocation lines. While three-dimensional imaging techniques such as atom-probe tomography (APT) [22] can measure the local variation of composition at the nanoscale, the identification of different phases and their crystal structures is beyond the scope of such a method. An indirect method that has been impactful in the literature uses APT to identify a second-phase by local composition variations and comparison with an equilibrium phase diagram [22]. At the same time, our atomistic simulations demonstrate a multi-phase coexistence in nanoscale precipitates that has not been reported to date, uncovering the complexity of possible linear complexion states that are available. For example, both $L1_0$ and $L1_2$ phases in nanoscale precipitates have an fcc-like structure but different compositions (see Fig. 3(b)), while both B2 and $L1_0$ phases have near-equiatomic compositions but different structures. In addition, metastable phases typically require additional theoretical investigations or first-principles calculations to determine their structural and thermodynamic properties such as lattice parameters, bulk energies, and interfacial energies. Our atomistic simulations and thermodynamic calculations demonstrate that metastable nanoscale precipitates can nucleate at dislocations if their bulk energy is lower than that for a corresponding solid solution and if they are able to form low-energy coherent interfaces with the matrix phase. There are likely past examples of phase transformations that were unknowingly caused by a linear complexion transition. For example, the experimental work of Nes [44] on precipitation of the cubic $Al_3Zr$ phase found that this phase nucleated near dislocations in subperitectic face-centered cubic Al-Zr alloys. The energetics of such a transformation and the early stages of nucleation require an atomic-scale perspective like that uncovered here.

In conclusion, we have discovered a novel phase transformation path at the nanoscale involving metastable phase formation and leading to a unique three-phase coexistence at the compression side of edge dislocations in Ni-doped Fe alloys. The nanoscale precipitates made of such phases form arrays along dislocation lines and can therefore be classified as linear complexions. Our atomistic simulations provide much more detail about the new phenomena of stable and metastable phase coexistence at the nanoscale than has previously been found in thermodynamic models or experiments. Using thermodynamic calculations, we determine that this phenomenon is caused by (1) an elevated composition in the dislocation segregation zone, (2) coherent interfaces between the metastable intermetallic and stable matrix phases, and (3) structural transition from metastable to stable phase accompanied by local strains. Because precipitation on dislocations will affect a wide range of mechanical properties of crystalline materials, this improved understanding of linear complexions will have broad scientific and technological impact.

**Acknowledgements:**

This research was supported by U.S. Department of Energy, Office of Basic Energy Sciences, Materials Science and Engineering Division under Award No. DE-SC0014232.

References:

[1]     S. J. Dillon, M. Tang, W. C. Carter, and M. P. Harmer, Acta Mater. **55**, 6208 (2007).

[2]     J. Luo, Appl. Phys. Lett. **95** (2009).

[3]     P. R. Cantwell, M. Tang, S. J. Dillon, J. Luo, G. S. Rohrer, and M. P. Harmer, Acta Mater. **62**, 1 (2014).

[4]     T. Frolov, D. L. Olmsted, M. Asta, and Y. Mishin, Nat. Commun. **4** (2013).

[5]     J. Luo and X. Shi, Appl. Phys. Lett. **92** (2008).

[6]     T. Frolov, M. Asta, and Y. Mishin, Phys. Rev. B **92**, 020103 (2015).

[7]     T. Frolov, M. Asta, and Y. Mishin, Curr. Opin. Solid State Mater. Sci. **20**, 308 (2016).

[8]     X. Shi and J. Luo, Phys. Rev. B **84** (2011).

[9]     D. Raabe, M. Herbig, S. Sandlöbes, Y. Li, D. Tytko, M. Kuzmina, D. Ponge, and P. P. Choi, Curr. Opin. Solid State Mater. Sci. **18**, 253 (2014).

[10]    N. Zhou and J. Luo, Acta Mater. **91**, 202 (2015).

[11]    T. Frolov, W. Setyawan, R. J. Kurtz, J. Marian, A. R. Oganov, R. E. Rudd, and Q. Zhu, Nanoscale **10**, 8253 (2018).

[12]    C. H. Liebscher, A. Stoffers, M. Alam, L. Lymperakis, O. Cojocaru-Mirédin, B. Gault, J. Neugebauer, G. Dehm, C. Scheu, and D. Raabe, Phys. Rev. Lett. **121**, 015702 (2018).

[13]    Q. Zhu, A. Samanta, B. Li, R. E. Rudd, and T. Frolov, Nat. Commun. **9** (2018).

[14]    S. J. Dillon, K. Tai, and S. Chen, Curr. Opin. Solid State Mater. Sci. **20**, 324 (2016).

[15]    T. J. Rupert, Curr. Opin. Solid State Mater. Sci. **20**, 257 (2016).

[16]    B. B. Straumal, A. A. Mazilkin, and B. Baretzky, Curr. Opin. Solid State Mater. Sci. **20**, 247 (2016).

[17]    T. Frolov, S. V. Divinski, M. Asta, and Y. Mishin, Phys. Rev. Lett. **110**, 1 (2013).

[18]    A. Khalajhedayati, Z. Pan, and T. J. Rupert, Nat. Commun. **7**, 10802 (2016).

[19]    G. Kim, X. Chai, L. Yu, X. Cheng, and D. S. Gianola, Scripta Mater. **123**, 113 (2016).

[20]    Z. Pan and T. J. Rupert, Acta Mater. **89**, 205 (2015).

[21]    V. Turlo and T. J. Rupert, Acta Mater. **151**, 100 (2018).

[22]    M. Kuzmina, M. Herbig, D. Ponge, S. Sandlöbes, and D. Raabe, Science **349** (2015).

[23]    A. Kwiatkowski da Silva, G. Leyson, M. Kuzmina, D. Ponge, M. Herbig, S. Sandlbes, B. Gault, J. Neugebauer, and D. Raabe, Acta Mater. **124**, 305 (2017).

[24]    P. Kontis, Z. Li, D. M. Collins, J. Cormier, D. Raabe, and B. Gault, Scripta Mater. **145**, 76 (2018).

[25]    A. Kwiatkowski Da Silva, D. Ponge, Z. Peng, G. Inden, Y. Lu, A. Breen, B. Gault, and D. Raabe, Nat. Commun. **9**, 1 (2018).

[26]    V. Turlo and T. J. Rupert, Scripta Mater. **154**, 25 (2018).

[27]    R. K. Koju, K. A. Darling, L. J. Kecskes, and Y. Mishin, JOM **68**, 1596 (2016).

[28]    T. O. Saetre, N. Ryum, and O. Hunderi, Mater. Sci. Eng. A **108**, 33 (1989).

[29]    T. Gladman, Mater. Sci. Technol. **15**, 30 (1999).

[30]    J. Y. He, H. Wang, H. L. Huang, X. D. Xu, M. W. Chen, Y. Wu, X. J. Liu, T. G. Nieh, K. An, and Z. P. Lu, Acta Mater. **102**, 187 (2016).

[31]    V. I. Levitas, Phys. Rev. B **70**, 1 (2004).

[32]    V. I. Levitas and M. Javanbakht, Nanoscale **6**, 162 (2014).

[33]    S. Y. Hu and L. Q. Chen, Acta Mater. **49**, 463 (2001).

[34]    F. Léonard and M. Haataja, Appl. Phys. Lett. **86**, 181909 (2005).

[35]    C. Hin, Y. Brechet, P. Maugis, and F. Soisson, Philos. Mag. **88**, 1555 (2008).

[36]     C. Hin, Y. Bréchet, P. Maugis, and F. Soisson, Acta Mater. **56**, 5535 (2008).

[37]     Z. Pan and T. J. Rupert, Phys. Rev. B **93**, 1 (2016).

[38]     S. Yang, N. Zhou, H. Zheng, S. P. Ong, and J. Luo, Phys. Rev. Lett. **120**, 85702 (2018).

[39]     S. Plimpton, J. Comput. Phys. **117**, 1 (1995).

[40]     G. Bonny, R. C. Pasianot, and L. Malerba, Modell. Simul. Mater. Sci. Eng. **17**, 025010 (2009).

[41]     B. Sadigh, P. Erhart, A. Stukowski, A. Caro, E. Martinez, and L. Zepeda-Ruiz, Phys. Rev. B **85**, 1 (2012).

[42]     A. Stukowski, Modell. Simul. Mater. Sci. Eng. **18**, 015012 (2010).

[43]     A. Stukowski, V. V. Bulatov, and A. Arsenlis, Modell. Simul. Mater. Sci. Eng. **20**, 085007 (2012).

[44]     E. Nes, Acta Metall. **20**, 499 (1972).

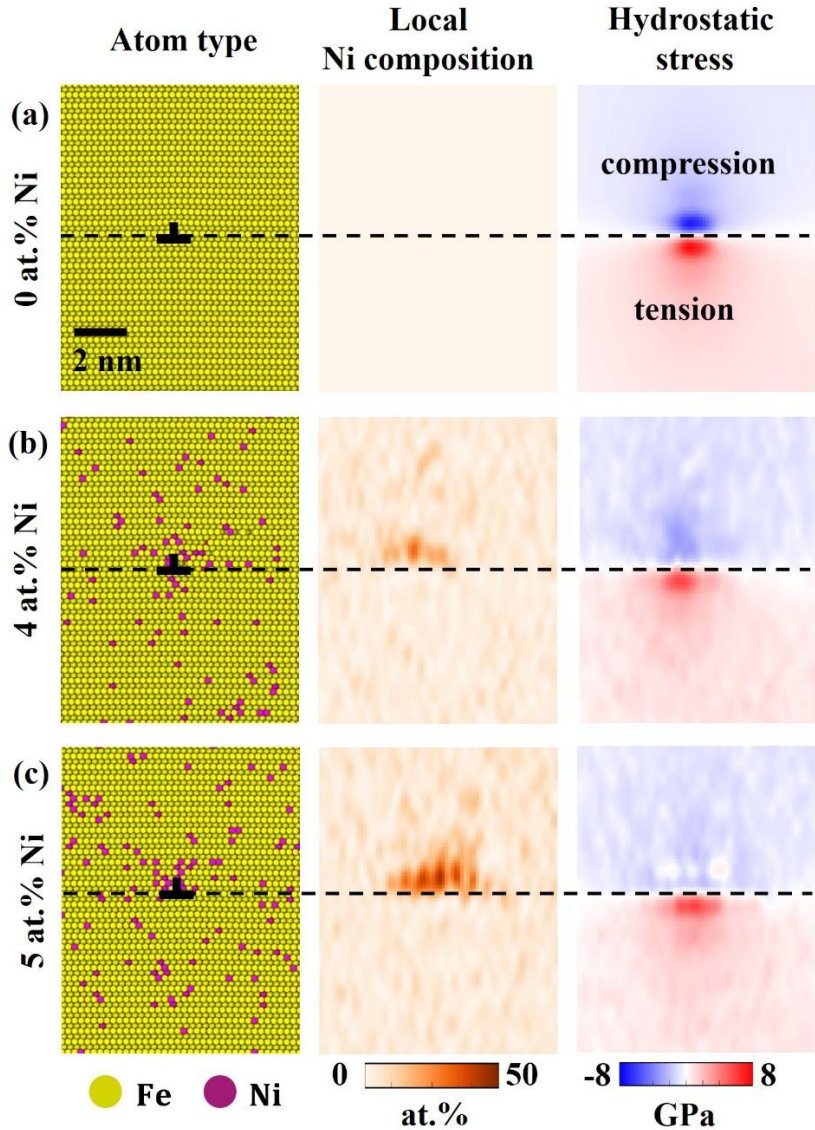

FIG. 1. Atomic snapshots, local composition, and local stress around one of the dislocation cores for the samples equilibrated at 600 K and having (a) 0 at.% Ni, (b) 4 at.% Ni, and (c) 5 at.%. Dashed lines represent the dislocation slip planes.

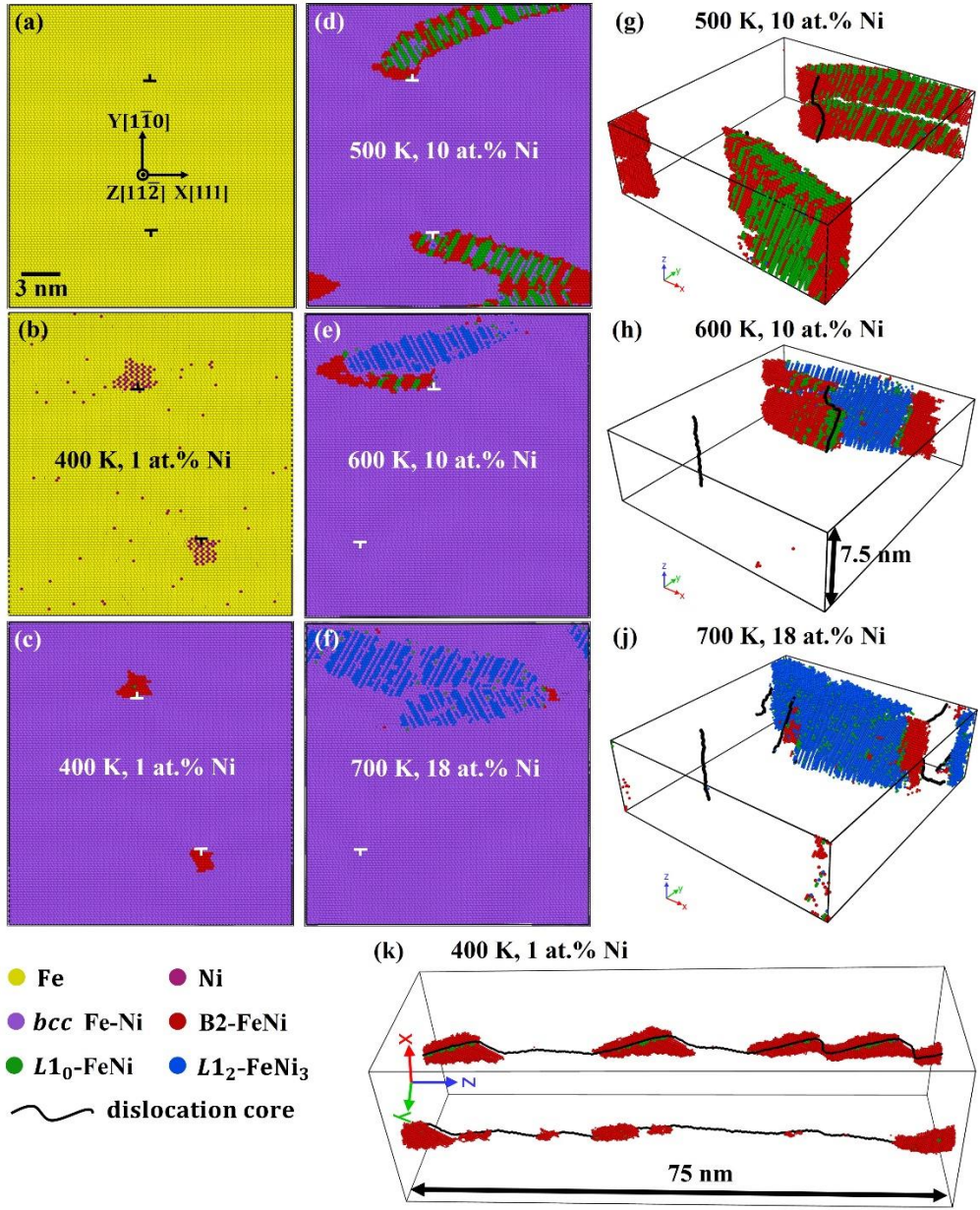

FIG. 2. View along the XY plane of the simulation cell with two dislocations (a) before and (b-f) after equilibration at the listed compositions and temperatures. Perspective views of the (d-f) samples are shown in (g-j). The long sample demonstrating precipitate arrays is shown in (k). Atoms in the bcc Fe-Ni solid solution are removed from (g-k).

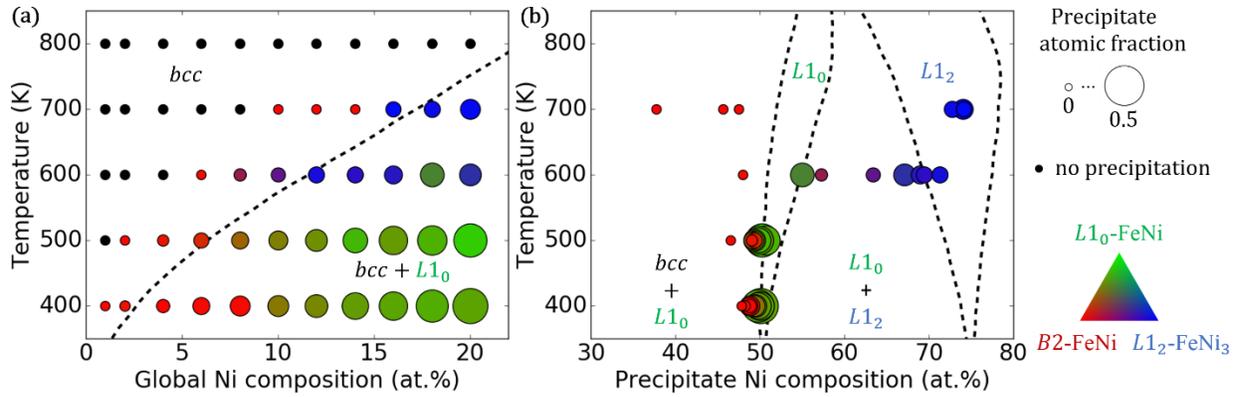

FIG. 3. (a) The complexion (phase) diagram, showing the size of the precipitates and the atomic fractions of the B2, L1$_0$, and L1$_2$ phases in each precipitate. (b) Local precipitate composition–temperature plot, in which the data correspond to the same samples shown in (a). Dashed lines represent the equilibrium phase diagram curves for the Fe-Ni interatomic potential used in this study [40].

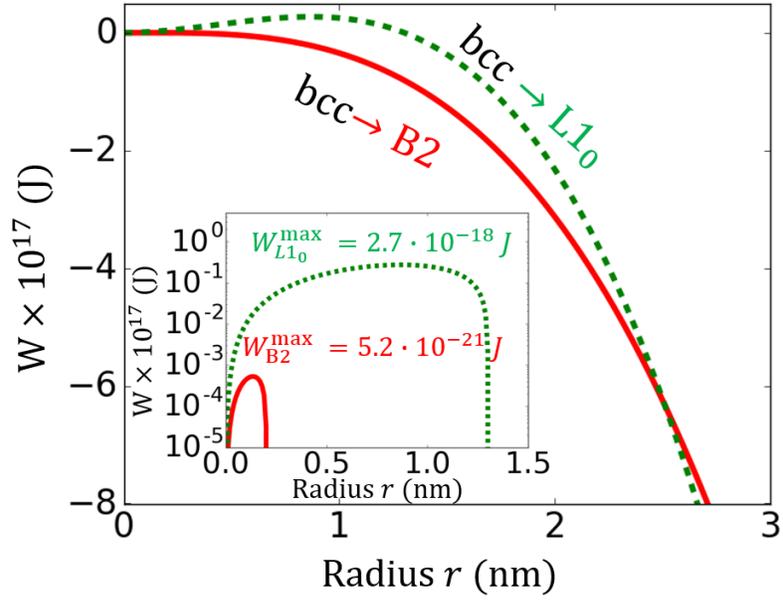

FIG. 4. The change in Gibbs free energy as a function of the radius of the intermetallic B2 (solid red line) and $L1_0$ (dashed green line) spherical precipitates formed in an equiatomic Fe-Ni solid solution. The corresponding values of the nucleation barriers (maximum values) are shown as well on the inset that shows the positive values of the main plot on a logarithmic scale.

**Intermetallic linear complexions: metastable phase formation and phase coexistence at dislocations**

Vladyslav Turlo [a], Timothy J. Rupert [a, b, *]

[a] Department of Mechanical and Aerospace Engineering, University of California, Irvine, CA 92697, USA

[b] Department of Chemical Engineering and Materials Science, University of California, Irvine, CA 92697, USA

## Estimation of nucleation barriers

To understand why a metastable B2 phase formation can be more favorable than an $L1_0$ phase at the nanoscale, we estimate the energy barriers for nucleation of spherical B2 and $L1_0$ precipitates in the equiatomic (50-50 at.%) Fe-Ni composition. This mimics phase transitions that occur in the regions of an elevated composition inside of the dislocation segregation zone (see Fig. S3). As shown in Fig. 1(a-c), the segregation of Ni to the compression side of the dislocation core quickly reduces the local compressive stresses. After reaching a near-equiatomic composition in the dislocation segregation zone (Fig. 1(c)), the local compressive stresses have become very close to zero. Thus, large external stresses from the dislocation are not present in the segregation zone and do not need to be treated when describing the formation of intermetallic phases within this region. However, as demonstrated in Fig. S2, the formation of intermetallic phases does lead to local tensile stresses inside of the dislocation segregation zone, and these types of stresses are considered in the thermodynamic treatment below. In classical nucleation theory, the work, $W$, associated with the formation of a spherical particle is expressed as [1]:

$$W = -\frac{4}{3}\pi r^3 \Delta g + 4\pi r^2 \gamma + \frac{4}{3}\pi r^3 \frac{\sigma \epsilon}{2} \qquad (1)$$

where $\Delta g = \Delta G / \overline{V}$ is a change in Gibbs free energy $G$ per volume $\overline{V}$, $\gamma$ is the interfacial energy between a solid solution and an intermetallic phase, $\sigma$ and $\epsilon$ are the hydrostatic stress and volumetric strain associated with the transformation, and $r$ is the precipitate radius. For the purpose of approximation, we also assume zero temperature and zero external pressure conditions, which simplifies the formulation for the Gibbs free energy as:

$$\Delta G = \Delta U = \Delta E_p \qquad (2)$$

or the difference in potential energies of solid solution and intermetallic phase. Thus,

$$\Delta g = \Delta E_p / \overline{V} = \Delta e_p / \overline{\Omega} \qquad (3)$$

where $e_p$ is potential energy per atom and $\overline{\Omega}$ is an average atomic volume. Considering transitions from the body-centered cubic (bcc) Fe-Ni solid solution to the B2 and to the L1$_0$ intermetallic phases, the corresponding $\Delta e_p$, $\overline{\Omega}$, $\sigma$, $\epsilon$, and $\gamma$ parameters are determined below.

To determine dependencies of the potential energy per atom on atomic volume for the three phases of interest, we considered a cubic simulation cell with sizes of $20a_0 \times 20a_0 \times 20a_0$, where $a_0$ is a lattice parameter. Lattice parameters were adjusted for each phase to values corresponding to atomic volumes in a range from 8 to 14 Å$^3$ with step of 0.1 Å$^3$. A curve for the bcc Fe-50%Ni solid solution was obtained by averaging over 100 different random distributions. Curves for the three phases are presented in Fig. S4, with and energy minimas extracted and listed in Table S1 together with the corresponding atomic volumes.

To calculate nucleation barriers for bcc to B2 and bcc to L1$_0$ phase transformations, we also determined average interfacial energies between the solid solution and intermetallic phases. To do so, we considered two of the most common coherent [110] interfaces between bcc and B2, and two of the most common incoherent $[110]_{bcc}/[111]_{fcc}$ interfaces between bcc and $L1_0$, as demonstrated in Fig. S5. The sizes $X \times Y \times Z$ of the simulation boxes shown in Fig. S4 were (a) $17 \times 12 \times 40$ nm$^3$, (b) $11.5 \times 12 \times 19$ nm$^3$, (c) $17 \times 24 \times 40$ nm$^3$, and (d) $11.5 \times 24 \times 19$ nm$^3$. Such relatively large sizes in X and Z directions allow for bcc/$L1_0$ lattice mismatches lower than 0.1%. Small shifts from 10% to 100% of a lattice period in each direction were applied in the X and Z directions to determine the lowest energy configurations for each bcc/ $L1_0$ interface. 100 random distributions of the bcc solid solution were tested and the resulting average interfacial energies were determined and shown in Fig. S5. The interfacial energies for the coherent bcc/B2 interfaces are around one order of magnitude lower than for the uncoherent bcc/$L1_0$ interfaces. Taking into account the relatively small differences in the bulk energies (the same order of magnitude), such a huge difference in interfacial energies already suggests a much higher nucleation barrier for the $L1_0$ phase as compared to the B2 phase.

However, second-phase formation in solids is also impacted by an internal strain energy penalty that is always positive and contributes to the nucleation barrier. While accurate expressions for the energy of inclusion in terms of volumetric strain and elastic constants are available [2], these treatments require one to know the elastic constants of each possible phase. Unfortunately, the determination of these elastic constants for a metastable phase is impossible with molecular dynamics, as any structural relaxation of the phase immediately leads to a complete or partial transformation to a more stable phase (see, for example, Fig. S6). As a result, it is not possible to precisely calculate the contribution of the elastic strain energy to the nucleation barrier of the B2 phase. To understand if this contribution may alter the preferential formation of the metastable phase, we estimate an extreme cases of a maximum possible strain energy for the B2 phase. The maximum strain energy can be obtained using the Eshelby inclusion theory [3].

According to this theory, the second-phase precipitate can be added in three steps: (1) removing some volume of the matrix phase, (2) filling this volume with the precipitate that has been elastically strained to fit the cavity, and (3) relaxing the obtained structure. In this case, the internal strain energy can be represented as $E_{strain} = \sigma\epsilon/2 - E_{relax}$, where $\epsilon$ and $\sigma$ are the volumetric strain and the corresponding hydrostatic stress, respectively, of the inclusion required to fit the cavity and $E_{relax}$ is the relaxation energy. Obtaining the accurate relaxation energy for most of the cases is extremely challenging, so we have used the conservative case and set this value to be equal to zero, giving the maximum possible strain energy associated with the B2 phase formation as $E_{strain}^{max} = \sigma\epsilon/2$. The volumetric strain associated with the bcc to B2 phase transformation can be estimated as $\epsilon_{bcc\rightarrow B2} = (\Omega_{bcc} - \Omega_{B2})/\Omega_{B2}$, where $\Omega_{bcc}$ and $\Omega_{B2}$, listed in Table S1, correspond to the equilibrium atomic volumes of the bcc solid solution and the B2 intermetallic phase, respectively. The hydrostatic stress, $\sigma_{bcc\rightarrow B2}$, was obtained by rescaling the cubic simulation cell of the B2 phase reflecting the volumetric strain, $\epsilon_{bcc\rightarrow B2}$. For the L1$_0$ phase, it is common to assume that incoherent interfaces lead to a plastic relaxation energy that is equal to the strain energy penalty from entering the precipitate [3]. It is important to note that this again is a conservative choice, as the existence of a non-zero elastic strain energy penalty for the L1$_0$ phase would only make this phase even more unfavorable. The stress and strain values corresponding to each phase transition are listed in Table S2. To estimate energy barriers for bcc solid solution transitions to either B2 or L1$_0$ intermetallic phases, we also determined the corresponding $\Delta e_p$, $\overline{\Omega}$, and $\gamma$ parameters (see Table S2) for the work associated with a formation of a spherical particle of each intermetallic phase, using the data listed in Table S1 and in Fig. S5. The critical radius, $r^{max}$, at which $W$ reaches the maximum value may be determined from the condition $\frac{dW}{dr} = 0$.

Consequently, the critical radius is $r^{\max} = 2\gamma \Big/ \left( \frac{\Delta e_p}{\bar{\Omega}} - \frac{\sigma\epsilon}{2} \right)$ and the corresponding energy barrier is

$W^{\max} = 16\pi\gamma^3 \Big/ 3\left( \frac{\Delta e_p}{\bar{\Omega}} - \frac{\sigma\epsilon}{2} \right)^2$. The energy barriers associated with the transitions from an equiatomic Fe-Ni solid solution to the B2 and $L1_0$ intermetallic phases are tabulated in Table S2. If the nucleation barrier value itself is assumed to be independent of temperature, the probability to form an overcritical nucleus can be estimated as $p = \exp\left( W^{\max} \Big/ k_B T \right)$. Fig. S7 presents the nucleation probability for the two considered phase transformations in linear and logarithmic scales.

[1] K. F. Kelton and A. L. Greer (2010), in *Pergamon Materials Series*, Pergamon.

[2] T. Mura (1987). *Micromechanics of Defects in Solids.* Martinus Nijhoff Publishers, Dordrecht.

[3] Balluffi, R. W., Allen, S. M., Carter, W. C., & Kemper, R. A. (2005). *Kinetics of materials*. J. Wiley & Sons.

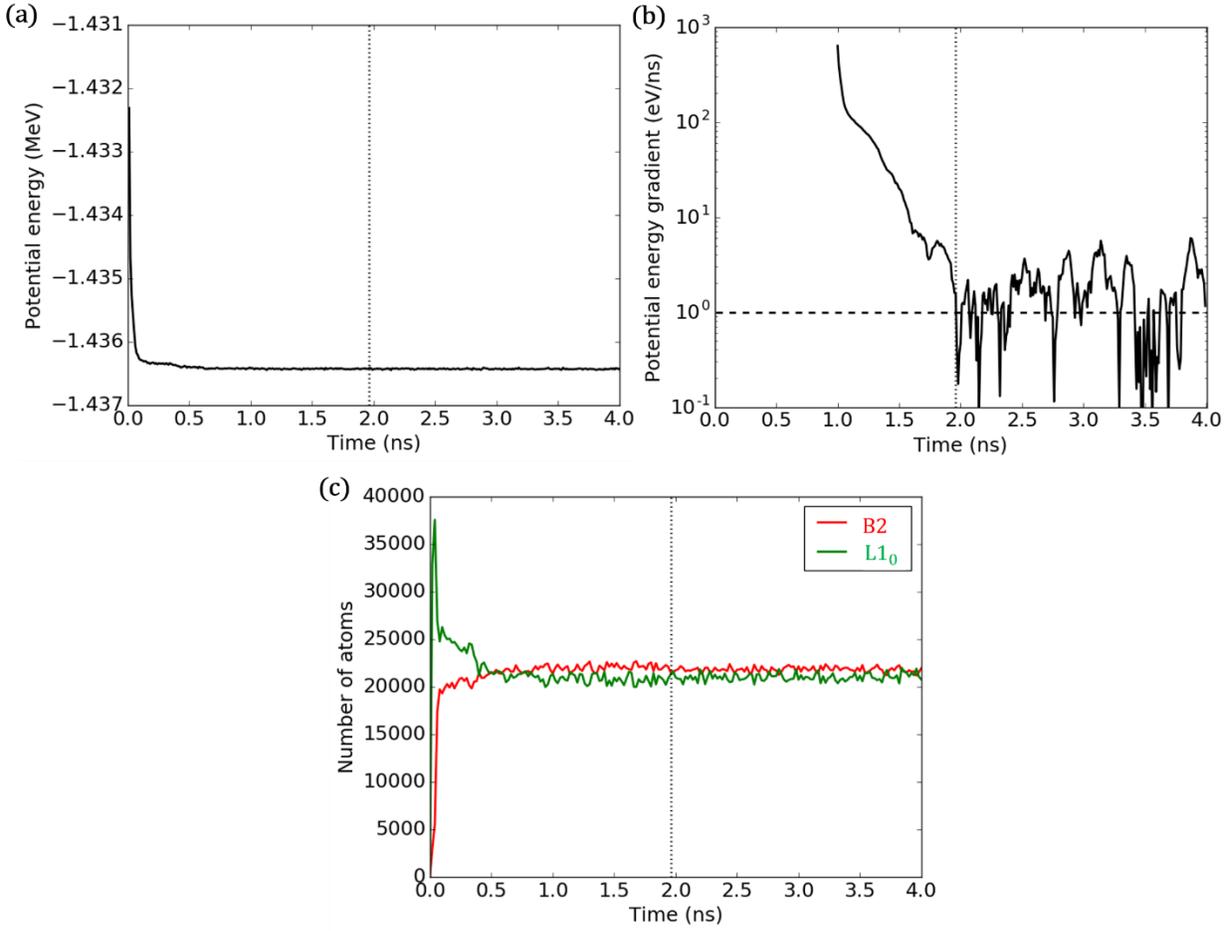

FIG. S1. Time evolution of (a) the potential energy, (b) the potential energy gradient, and (c) the number of atoms of the B2 and $L1_0$ phases for the sample with 10 at.% Ni equilibrated at 500 K. The horizontal dashed line in (b) represents the critical potential energy gradient, while the vertical dotted line shows the time at which critical potential energy gradient was reached for the first time.

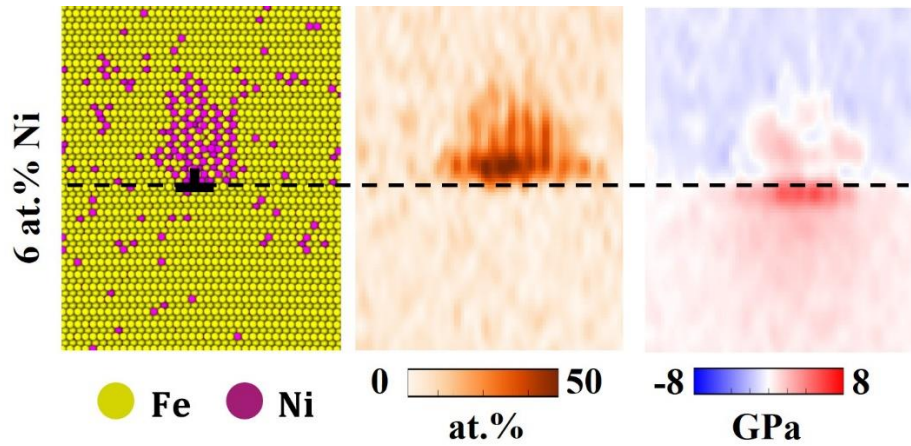

Fig. S2. An atomic snapshot, local composition, and local stress around one of the dislocation core for the sample equilibrated at 600 K and having 6 at.% Ni. The dashed line represent the dislocation slip planes.

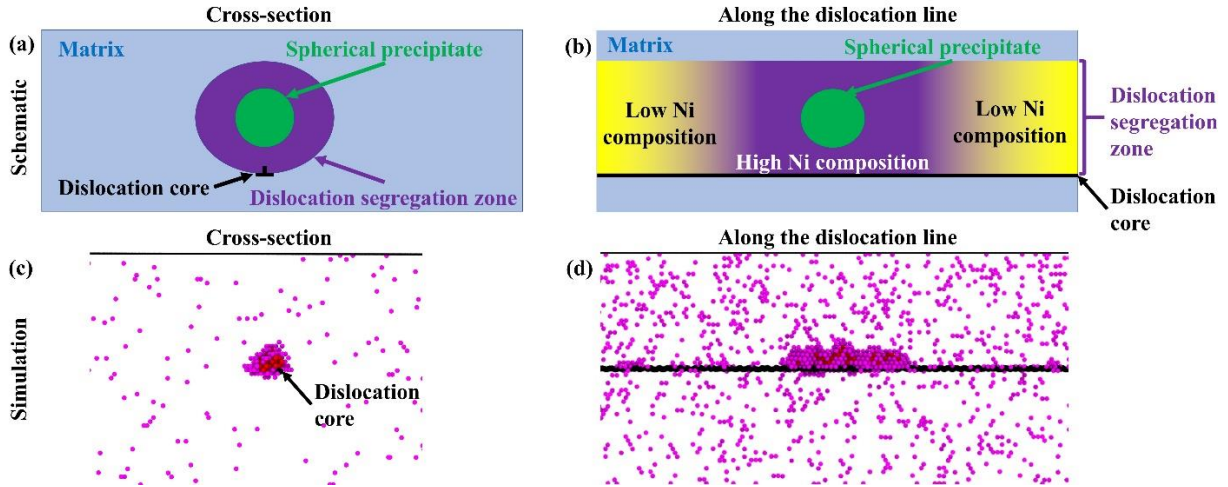

FIG. S3. (a,b) Schematic demonstrating the formation of a spherical precipitate of an intermetallic phase inside of the Ni-enriched zone of the dislocation segregation region that is experiencing spinodal decomposition. (c,d) A local equilibrium structure of the Fe-1at.%Ni sample annealed at 400 K (See Fig. 2(k)) demonstrating a Ni-rich zone in the dislocation segregation zone with the B2 phase precipitation. Only Ni atoms are shown. Black line represents the position of the dislocation core obtained by the DXA method.

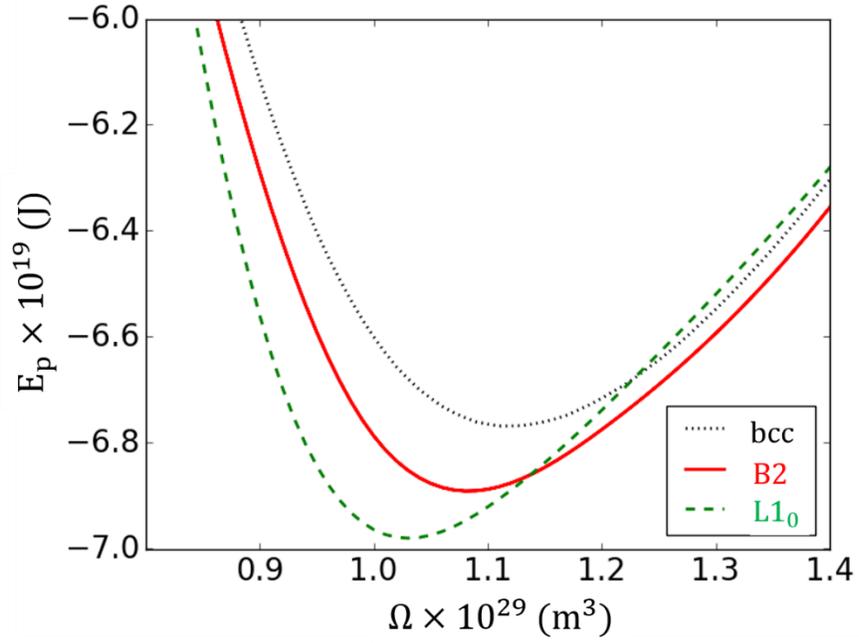

FIG. S4. The potential energy per atom as a function of the atomic volume for the three phases of interest at 0 K.

Table S1. Equilibrium potential energy per atom and the corresponding atomic volume for the three phases of interest.

| Phase \ Property | Potential energy per atom $e_p$, $J$ | Atomic volume $\Omega$, $m^3$ |
|---|---|---|
| Fe-50 at.% Ni | -6.768e-19 | 1.11e-29 |
| B2-FeNi | -6.891e-19 | 1.08e-29 |
| L1$_0$-FeNi | -6.980e-19 | 1.03e-29 |

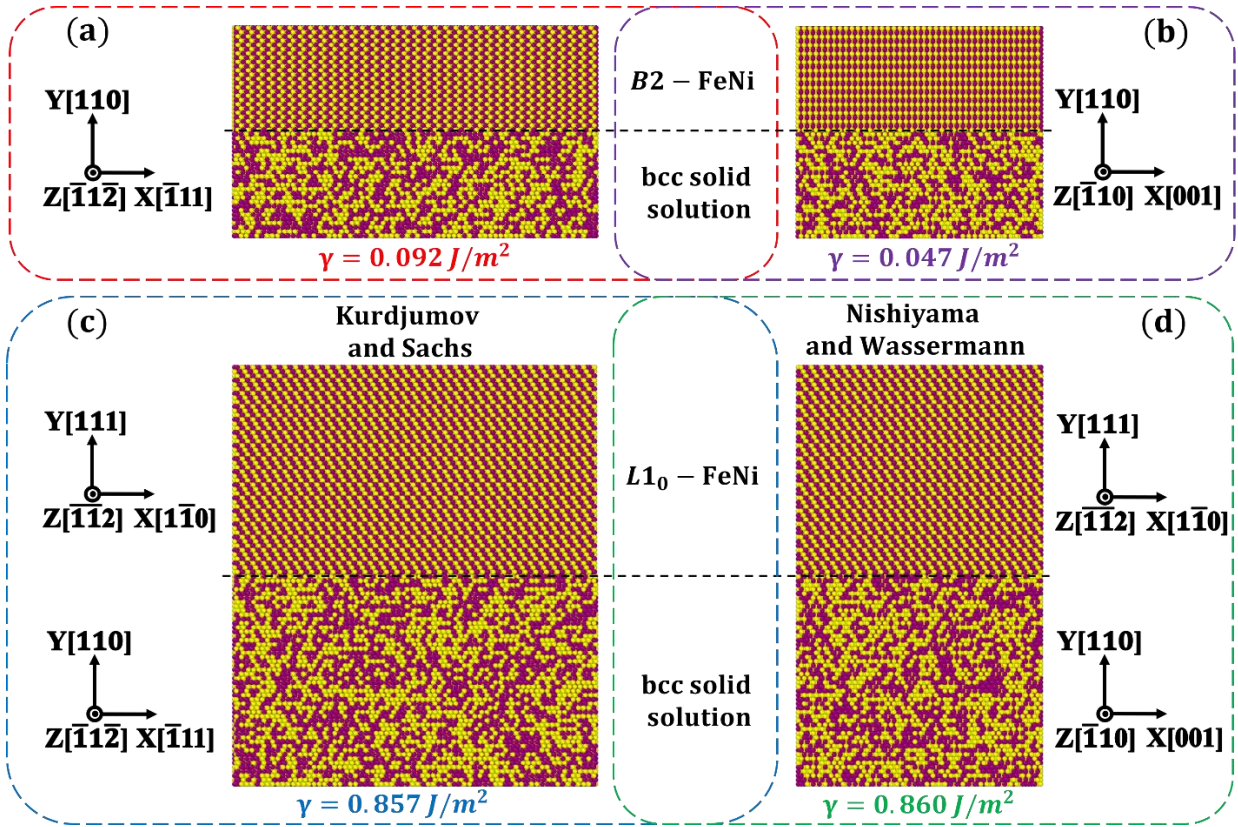

FIG.S5. Samples used to determine the interfacial energies between Fe-Ni solid solutions and B2 and L1$_0$ intermetallic phases are shown together with the corresponding values of these average interfacial energies.

Table S2. Model parameters, critical radii, and nucleation barriers for the two transitions from equiatomic bcc Fe-Ni solid solution to the B2 and L1$_0$ intermetallic phases.

| Transition | $\Delta e_p$, J | $\overline{\Omega}$, m$^3$ | $\gamma$, J/m$^2$ | $\epsilon$ | $\sigma$, Pa | $r^{\max}$, m | $\frac{W^{\max}}{k_B}$, K |
|---|---|---|---|---|---|---|---|
| **bcc→B2** | 0.123e-19 | 1.095e-29 | 0.0695 | 0.0278 | 6.05e9 | 1.34e-10 | 377 |
| **bcc→L1$_0$** | 0.212e-19 | 1.07e-29 | 0.8585 | 0 | 0 | 8.67e-10 | 195,661 |

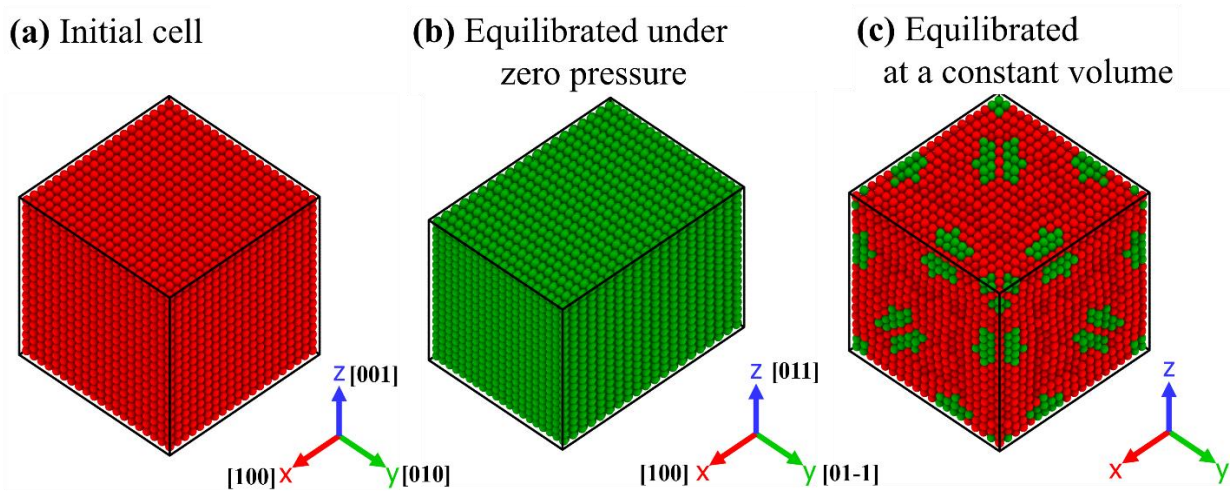

FIG. S6. The cubic cell of the B2 phase (a) before and (b,c) after equilibration using the conjugate gradient method (b) under zero external pressure and (c) at a fixed sample volume. In (c), the L1$_0$ phase starts to nucleate, but retains a B2 border to accommodate the local strains. The initial cubic simulation cell has a size of $20a_0 \times 20a_0 \times 20a_0$, where $a_0$ is a lattice parameter corresponding to the energy minima of the B2 phase (see Fig. S4). Atoms are colored according to an alloy type (red – B2, green – L1$_0$).

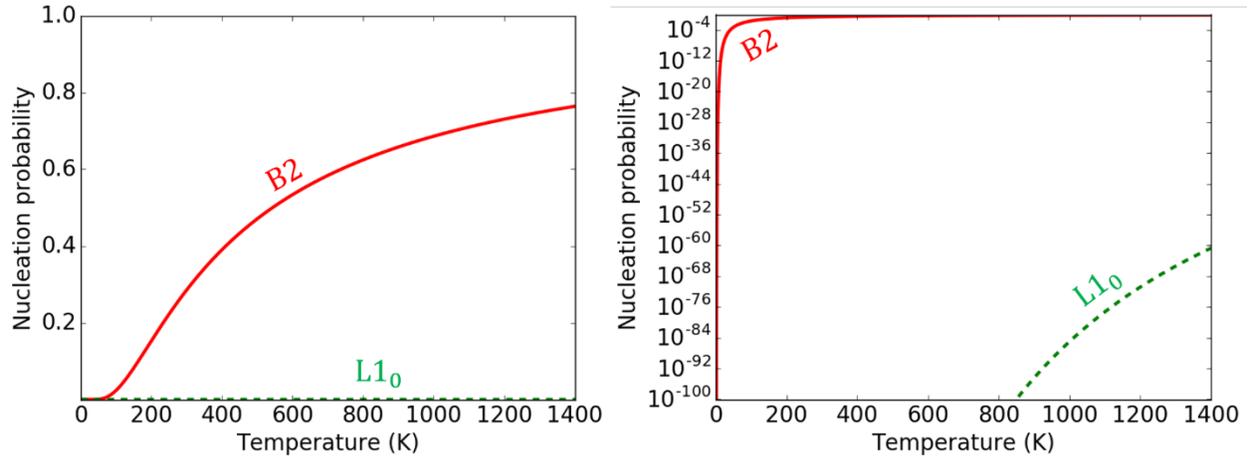

FIG.S7. The probability of forming a nucleus larger than the critical size for B2 (red solid line) and L1$_0$ (green dashed line) phases as a function of temperature, as determined by $p = \exp\left(W^{\max}/k_B T\right)$ and shown in linear (left) and logarithmic (right) scales.

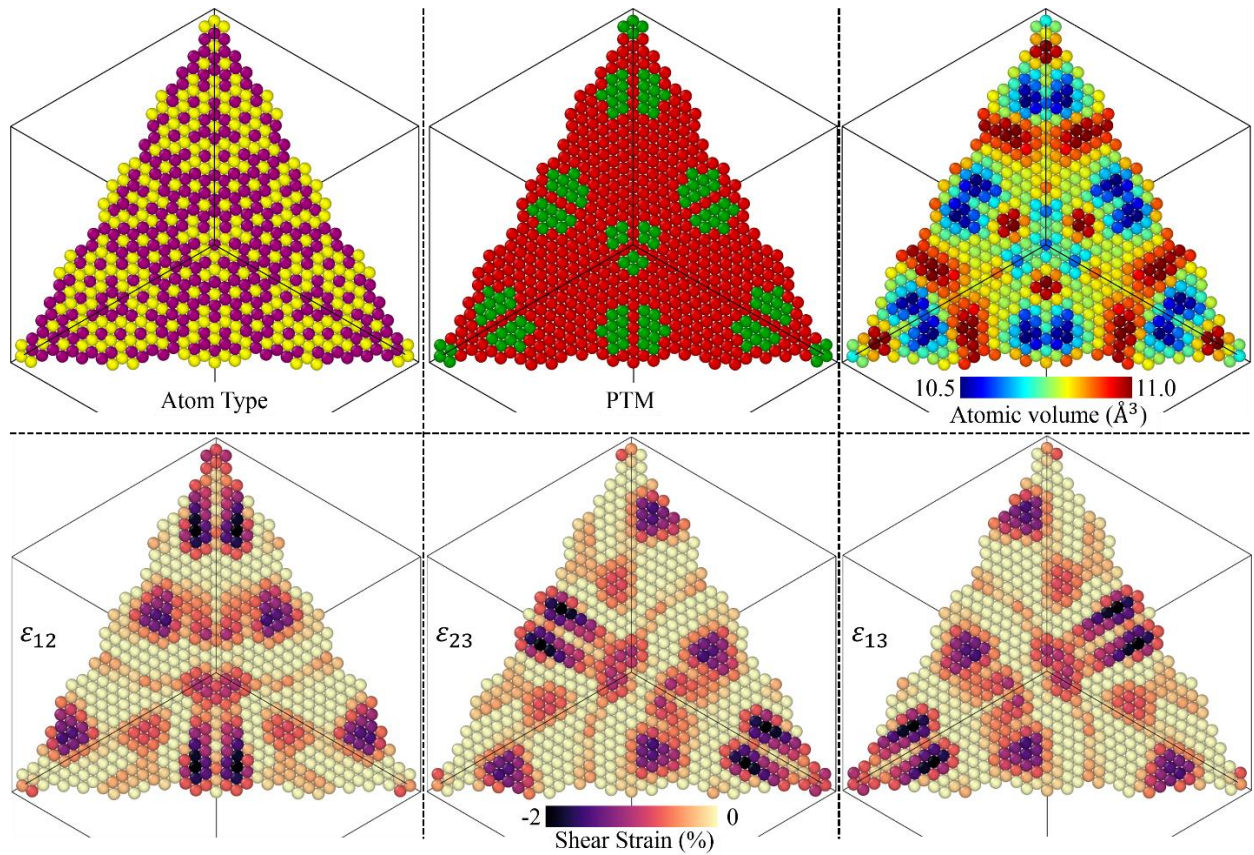

FIG. S8. The [111] plane of a cubic sample of the B2 phase equilibrated using the conjugate gradient method with a fixed sample volume (see Fig. S6(c)). The $L1_0$ phase starts to nucleate, but retains a B2 border to accommodate the local strains. The cubic simulation cell has a size of $20a_0 \times 20a_0 \times 20a_0$, where $a_0$ is a lattice parameter corresponding to the energy minima of the B2 phase (see Figure S4). Atoms are colored according to an atom type (yellow – Fe, magenta – Ni), an alloy type (red – B2, green – $L1_0$), atomic volume and shear strain (color bars are presented in the figure).